%% file: manuscript.tex
\documentclass[sigconf]{acmart}
\AtBeginDocument{%
  }

\setcopyright{acmlicensed}
\copyrightyear{2025}
\acmYear{2025}
\acmDOI{XXXXXXX.XXXXXXX}
\setcopyright{none}
\acmConference[Conference acronym 'XX]{Make sure to enter the correct
  conference title from your rights confirmation email}{June 03--05,
  2018}{Woodstock, NY}
\acmISBN{978-1-4503-XXXX-X/2018/06}

\usepackage{listings}
\usepackage{tabularx}
\usepackage{enumitem}

\copyrightyear{2025}
\acmYear{2025}
\setcopyright{rightsretained}
\acmConference[CHI EA '25]{Extended Abstracts of the CHI Conference on Human Factors in Computing Systems}{April 26-May 1, 2025}{Yokohama, Japan}
\acmBooktitle{Extended Abstracts of the CHI Conference on Human Factors in Computing Systems (CHI EA '25), April 26-May 1, 2025, Yokohama, Japan}\acmDOI{10.1145/3706599.3720137}
\acmISBN{979-8-4007-1395-8/2025/04}

\begin{document}

\title{Exploring Socio-Cultural Challenges and Opportunities in Designing Mental Health Chatbots for Adolescents in India}

\author{Neil K. R. Sehgal}
\email{nsehgal@seas.upenn.edu}
\affiliation{%
  \institution{University of Pennsylvania}
  \city{Philadelphia}
  \state{Pennsylvania}
  \country{USA}
}

\author{Hita Kambhamettu}
\email{hitakam@seas.upenn.edu}
\affiliation{%
  \institution{University of Pennsylvania}
  \city{Philadelphia}
  \state{Pennsylvania}
  \country{USA}
}

\author{Sai Preethi Matam}
\email{saipreethimatam@gmail.com}
\affiliation{%
  \institution{Mamata Academy  of Medical Sciences}
  \city{Hyderabad}
  \state{Telengana}
  \country{India}
}

\author{Lyle Ungar}
\email{ungar@cis.upenn.edu}
\affiliation{%
  \institution{University of Pennsylvania}
  \city{Philadelphia}
  \state{Pennsylvania}
  \country{USA}
}

\author{Sharath Chandra Guntuku}
\email{sharathg@seas.upenn.edu}
\affiliation{%
  \institution{University of Pennsylvania}
  \city{Philadelphia}
  \state{Pennsylvania}
  \country{USA}
}

\renewcommand{\shortauthors}{Trovato et al.}

\begin{abstract}

Mental health challenges among Indian adolescents are shaped by unique cultural and systemic barriers, including high social stigma and limited professional support. Through a mixed-methods study involving a survey of 278 adolescents and follow-up interviews with 12 participants, we explore how adolescents perceive mental health challenges and interact with digital tools. Quantitative results highlight low self-stigma but significant social stigma, a preference for text over voice interactions, and low utilization of mental health apps but high smartphone access. Our qualitative findings reveal that while adolescents value privacy, emotional support, and localized content in mental health tools, existing chatbots lack personalization and cultural relevance.  These findings inform recommendations for culturally sensitive chatbot design that prioritizes anonymity, tailored support, and localized resources to better meet the needs of adolescents in India. This work advances culturally sensitive chatbot design by centering underrepresented populations, addressing critical gaps in accessibility and support for adolescents in India.
\end{abstract}

\ccsdesc[500]{Human-centered computing~Interaction design process and methods}
\ccsdesc[500]{Human-centered computing~User centered design}

\keywords{mental health, user-centered design, India, adolescents, chatbots, conversational agents}


\maketitle

\section{Introduction}
Mental health challenges among adolescents are a global concern, with stress, anxiety, and depression emerging as significant issues during this transitional life stage \cite{patel2007mental}. In low- and middle-income countries like India, home to 20\% of the world's  under 25 population, these challenges are exacerbated by deeply systemic and societal barriers \cite{Naveed2020PrevalenceOC, gallup2025}. In India, 95\% of adults lack access to mental healthcare, the highest treatment gap across Asia \cite{murthy2017national, Naveed2020PrevalenceOC}.

Prior research has highlighted the potential of AI-driven mental health tools to provide scalable and anonymous support, yet much of this work has been developed using datasets from high income Western populations \cite{kozelka2021advancing, cho-etal-2023-integrative}. Studies have emphasized that cultural factors, such as language norms, emotion regulation strategies, and stigma, shape mental health experiences and require culturally tailored interventions \cite{rai2024key, krendl2020countries}. For instance, one meta-analysis found culturally adapted mental health services were up to four times more effective than services designed for broad audiences \cite{griner2006culturally}. As such, recent work in HCI has focused on characterizing the mental health landscape and designing digital mental health tools for various subgroups across South Asia including Indian working class women, Kashmiri young adults, and Bangladeshi adolescents \cite{reen2022improving, wani2024unrest, rahman2021adolescentbot}. Our work builds on these findings by focusing on the unique barriers and preferences of adolescents in India to inform the design of culturally inclusive mental health chatbots.

Through a mixed-method approach that involved a survey of 278 adolescents and qualitative interviews with 12 participants, we investigate three key questions. 

\begin{description}
    \item[RQ1:] \textbf{How do Indian adolescents perceive and navigate mental health challenges within societal and cultural contexts?}
    \item[RQ2:] \textbf{What are the experiences and challenges Indian adolescents face when using current digital tools to seek mental health support?}
    \item[RQ3:] \textbf{How can chatbots be designed to better support the mental health needs of Indian adolescents?}
\end{description}

We find that stigma is a key barrier to seeking formal mental health support, driving adolescents toward alternative strategies such as self-reliance, peer support, and digital resources. 
Our study finds that chatbots, in particular, present unique opportunities to fill this gap. Unlike traditional online search platforms, which often overwhelm users with uncurated and impersonal information, chatbots can deliver emotionally resonant, personalized interactions tailored to users' immediate needs. 
However, existing digital tools fall short due to issues of cultural irrelevance, lack of personalization, and insufficient privacy features. Based on these insights, we propose design recommendations to develop inclusive and culturally sensitive chatbot tools that address the needs of Indian adolescents, emphasizing features such as anonymity, emotional support, and locally relevant resources.

By addressing these challenges, this research contributes to the design of digital mental health interventions that are not only effective but also culturally aligned, ultimately enhancing accessibility and engagement for Indian adolescents navigating complex social and systemic landscapes.

\section{Methods}
\textbf{Study Design}
This study employed a mixed-methods approach to explore the mental health needs, barriers, and preferences of adolescents in India. The research was conducted in two phases: (1) a quantitative survey to capture broad trends and (2) in-depth qualitative interviews to gather rich, contextual insights. The survey provided a high-level understanding of participants’ experiences and attitudes, while the interviews allowed us to dive deeper into the nuances behind the quantitative findings. This approach allowed us to triangulate findings and develop comprehensive recommendations for the design of culturally inclusive digital mental health tools.

\textbf{Participant Recruitment}
Participants were recruited through a combination of online platforms, including social media and outreach to school administrators. We aimed to target participants from a range of geographic regions and socioeconomic backgrounds. Upon completion of the anonymous surveys, participants were redirected to a second survey, asking if they were interested in participating in an in-depth interview. All participants completing the survey were entered into a raffle for one of ten 2000 INR gift cards, and all interview participants received a 500 INR gift card.

\textbf{Phase 1: Survey}
The survey was designed to collect quantitative data on participants’ mental health experiences, perceived barriers to seeking support, and preferences for digital mental health tools. The survey consisted of both closed-ended questions (e.g., Likert scales and multiple-choice) and open-text fields for additional comments. Questions were developed with the assistance of mental health counselors in India and were also drawn from previously validated surveys and scales \cite{vogel2006measuring, murthy2017national, mental_health_survey_2022, lowe2005detecting}. Responses were collected via Qualtrics and all questions were in English.

\textbf{Phase 2: Qualitative Interviews}
To gain deeper insights into the themes identified in the survey, we conducted semi-structured interviews with 12 participants. The interviews primarily consisted of a role-play activity, along with in-depth questions about participants’ mental health experiences and preferences. 

The role-play activity was designed to simulate real-life challenges commonly faced by adolescents in India, based on discussions with mental health counselors. Participants were asked to choose one of six scenarios (see Appendix) reflecting these challenges and were asked to engage in two tasks:

\begin{itemize}
    \item \textbf{Searching for Information Online}: Participants were asked to search for information or solutions to the scenario using any online platform they felt comfortable with (e.g., Google, YouTube Shorts, Instagram). They were instructed not to use chatbots for this task. Participants shared their screens during this activity, allowing us to observe their search behavior and the resources they consulted.

    \item \textbf{Interacting with a Custom Chatbot}: Participants then engaged with a basic chatbot prototype we developed for the study (Appendix Figure 1). The chatbot was built on top of GPT-4o-mini and designed to be a supportive and empathetic conversational agent aimed at helping Indian students facing challenges. It was not intended to provide therapy or replace professional counselors but to offer understanding, encouragement, and practical suggestions. The chatbot’s responses were structured around a predefined prompt developed in collaboration with mental health counselors in India. The full prompt is provided in the Appendix.

\end{itemize}

Interviews were conducted via Zoom in English, with each session lasting approximately 45-60 minutes. This combination of role-play and qualitative inquiry provided rich, contextual insights into participants’ mental health needs, barriers, and preferences for digital mental health tools.

\textbf{Data Analysis}
Survey data were analyzed using descriptive statistics. Thematic analysis was employed to examine the interview transcripts. One author initiated the process by performing an open-coding pass to generate an initial set of codes. These codes were subsequently reviewed by a second author, who collaborated with the first author to refine them. This refinement involved discarding codes that were not instrumental in capturing participants' needs or proposing enhancements to digital mental health tools. To ensure validity, detailed discussions were conducted throughout each phase of the analysis, emphasizing a collaborative review to maintain accuracy and consistency in coding. This method was preferred over calculating inter-rater reliability (IRR), as the intricate nature of the codes and the context-sensitive discussions provided deeper insights \cite{mcdonald2019reliability, kambhamettu2024explainable}.

\textbf{Ethics}
The study was approved by our institutional review board, and all participants provided informed consent prior to participation. Survey responses were anonymous, and interviews were recorded with explicit consent.

\section{Phase 1: Survey Findings}

The survey included responses from 278 adolescents across India. Notable findings include:

\textbf{Limited Uptake:} A significant majority (66.7\%) had never accessed mental health services. Of those who sought help, 31.4\% consulted school counselors, while only 8.6\% used mental health apps, highlighting a large treatment gap.

\textbf{Low Self-Stigma but High Social Stigma: }Only 11.2\% of participants reported feelings of inadequacy or inferiority when considering therapy (self-stigma). However, 41.7\% expressed concerns about negative societal perceptions if they sought professional help (social stigma), underscoring the cultural barriers to formal care.

\textbf{Mental Health Chatbots: Low Usage but High Perceived Helpfulness:} While only 25\% had used chatbots to talk about their feelings or problems, 77\% of these users reported finding them helpful. This highlights their potential, despite low adoption rates, to address gaps in traditional care.

\textbf{Skewed Preferences for Text-Based Interactions:} A majority (69\%) preferred text-based communication with chatbots, compared to only 7\% favoring voice.

A summary of key survey results is provided in Table~\ref{tab:survey_data}.

\begin{table}[ht]
\caption{Survey Data Summary}
\label{tab:survey_data}
\centering
\small
\begin{tabular}{p{0.35\linewidth} p{0.6\linewidth}}
\toprule
\textbf{Characteristic} & \textbf{Details} \\
\midrule
\multicolumn{2}{l}{\textbf{Demographic Characteristics}} \\
Participants  & 278 students (60\% first-year, 35\% second-year) \\
Age           & 14--23 years (mean: 19 years, under 18: 12\% ) \\
Gender        & 74\% female, 22\% male, 3\% undisclosed \\
Religion      & 88\% Hindu \\
\midrule
\multicolumn{2}{l}{\textbf{Mental Health Challenges}} \\
Loneliness    & Over the past 12 months, 40\% felt lonely "sometimes", 26\% "most of the time" \\
Prolonged sadness &  Over the past 12 months, 36\% experienced sadness/hopelessness for >2 weeks \\
Sleep disruption  & Over the past 12 months, 36\% reported stress-related sleep disruption "occasionally", 16\% "most of the time" \\
Pressure for good grades & 34\% felt "a lot", 45.1\% felt "some" pressure \\
Social pressure to fit in  & 25\% experienced "a lot", 43.7\% "some" pressure \\
\midrule
\multicolumn{2}{l}{\textbf{Past Experience with Mental Health Support}} \\
Never accessed services & 66.7\% \\
Common resources used by those who had accessed services & Peer support groups (35.7\%), school counselors (31.4\%), private therapists (25.7\%), online counseling (11.4\%), mental health apps (8.6\%) \\
\midrule
\multicolumn{2}{l}{\textbf{Attitudes Towards Mental Health}} \\
Supportive communities & 58\% community, 65\% parents, 72\% friends deemed supportive/very supportive \\
Mental health discussions & Rarely/never with family (61\%); frequently/occasionally with friends (63\%) \\
\midrule
\multicolumn{2}{l}{\textbf{Barriers to Mental Health Support}} \\
Self-stigma   & If they sought therapy, 11.2\% would feel inadequate, 12.6\% self-dissatisfied, 10.5\% inferior \\
Social stigma & If they sought therapy, 41.7\% concerned about negative reactions from others \\
\midrule
\multicolumn{2}{l}{\textbf{Digital Resource Use and Preferences}} \\
Technology access & 87\% owned personal smartphones, 96\% frequently used the internet \\
Chatbot usage     & 81\% had used chatbots; 25\% for personal issues (77\% found these personal conversations helpful) \\
Desired features  & Anonymity (60\%), quick responses (58\%), access to resources (46\%), peer support (28\%) \\
\midrule
\multicolumn{2}{l}{\textbf{Chatbot Concerns and Desired Improvements}} \\
Key concerns      & Privacy (67\%), lack of personal connection (47\%), cost (42\%), reliability (32\%) \\
Desired features  & Coping strategies (72\%), personalized recommendations (70\%), regular check-ins (50\%) \\
Preferred communication & Text (69\%), voice (7\%), both (23\%) \\
Language preference     & 94\% English \\
\bottomrule
\end{tabular}
\end{table}

\section{Phase 2: Interview Findings}

We interviewed 12 participants ranging in age from 14 to 20. The majority (83\%) of participants identified as female, reflecting the makeup of survey participants. A plurality (n=5) of participants chose to roleplay the scenario of struggling to get into university. Appendix Table 1 displays a breakdown of participant demographics.

\subsection{How Indian Adolescents Perceive and Navigate Mental Health}


\subsubsection{Stigma and fear of judgement} \label{stigma}
As in the survey, many participants (N=9) cited social stigma as a major barrier to seeking mental health support. Notably, while a majority of survey respondents cited supportive attitudes from their family, most survey respondents reported rarely discussing mental health with family. Similarly, no interviewee described being comfortable discussing their mental health with their parents. P9 explained, “I wouldn’t seek official mental health support because I don’t want it on record—I know my parents wouldn’t be okay with that.” This highlights the need for digital tools that provide anonymity, avoid excessive data collection, and give users control over what information is shared or stored. Some participants also expressed reluctance to share their struggles even with close friends. Generational differences also shaped perceptions, with older family members frequently viewing mental health as an indulgence or weakness. P12 shared, “They think it’s just another way to extort money because they never had these issues in their time.” Some participants expressed the unique difficulties of growing up in rural areas. Reflecting on her rural upbringing, P5 shared, “Mental health problems are not considered as problems. It’s just overthinking.”

\subsubsection{Additional Barriers} \label{barriers}
In line with our survey findings on low uptake of counseling, only one participant had sought counseling in the past, and few knew others who had. Apart from stigma, participants cited not knowing who to reach out to. P5 recounted, “I can recall my days of my 10th standard…struggling with so many things, and I was like, whom do I reach out? And I had no one to reach out to.” P5 additionally found little support from teachers. “They were like you are just taking stress because of your boards and everything will be sorted,” she shared. Other participants suggested cost barriers. P1 stated, “it might even be quite costly for a student to go and consult a therapist.” One participant, who had previously sought counseling, was not fully satisfied. P10 explained “When I go to counselors sometimes they're not able to understand my situation.”

\subsubsection{Reliance on friends for support} \label{friends}
Our survey found high levels of supportive attitudes from friends. Similarly, interviewees cited friends as their most common source (N=6) of emotional support, offering a judgment-free space for discussing mental health concerns. P1 shared, “Talking to friends is how we usually deal with mental health problems. I don’t know anyone who has consulted a therapist.” However, participants acknowledged that not all of their friends felt comfortable sharing their problems, and others mentioned limitations of relying solely on friends. P1 recounted feeling hesitant towards going to their friends again after a recent issue they shared, stating “I felt something like being judged.” This suggests a gap where digital tools could complement informal peer networks by providing a judgment-free, always-available source of support.

\subsubsection{Self-reliance and digital resources}
When traditional support systems were unavailable, participants turned to self-directed strategies, including hobbies, sleeping, journaling, and online platforms. P3 noted, “If I have no one near me, I just take my phone, open Google, and type my problem to seek answers.” However, many expressed dissatisfaction with existing digital resources, citing issues of reliability, lack of personalization, and overwhelming amounts of information. 

\subsection{Experiences with Current Technologies Around Mental Health}

\subsubsection{Navigating Search Engines}
Google was a common starting point for participants seeking mental health advice due to its accessibility and familiarity. However, most participants found it challenging to navigate the volume of results or determine their reliability. P11 noted, “I had to search a bit after typing the topic. It wasn’t easy to find what I wanted.” P12 stated, "I just want a gist of what I need to know, not long articles." Search engines and websites were criticized for providing excessive, uncurated information or hiding relevant details behind paywalls. P4 remarked, "It’s all over the place—there are no curated spaces for specific mental health information." Two participants also highlighted that navigating the results while anxious or depressed would be too difficult. P4 stated, “if I were to experience this scenario I wouldn't be as critical as I am right now…I wouldn't be very proactive in my search. I would already be anxious, and it would be a lot of work to go through reliable sources.”

\subsubsection{Lack of Relevance} \label{relevance}
The lack of culturally relevant, trustworthy mental health information was a recurring frustration for participants. P6 shared, “It was a bit helpful but not completely helpful for us.” P9 stated, “I found a helpline, but it was for California—how is that helpful to me?” P4 explained, “As an Indian, I'm experiencing something extremely different from something an American or an Australian is experiencing when it comes to not just exams, but also things to do with relationships and financial constraints”. This highlights the need for digital tools that curate locally relevant information and connect users to appropriate resources.

\subsubsection{Role of Social Media and Community Platforms}
While the majority of survey respondents mentioned an interest in self-help tips, fewer interviewees shared this. Instead, many interviewees discussed wanting to read about others' personal experiences. P3 emphasized, "I want reassurance that someone went through this and succeeded." Still some disagreed. P4 commented, "There’s very little that personal testimonies alone can do to help." Moreover, while platforms like Quora and Reddit provided opportunities to seek advice and share experiences, they were often viewed as not specific enough. P3 shared, “Quora feels less trustworthy because everyone’s experience is different.”

\subsubsection{Existing Chatbots}
All participants had used chatbots for school, but only four had used them for personal or mental health conversations, mirroring the survey distribution. Those who had personal conversations with chatbots valued them for accessibility and anonymity, seeing them as viable alternatives when feeling judged by others. P3 explained, “When I feel judged by my friends or family, I’d rather use a chatbot.” P5 shared, “I used ChatGPT to calm myself when I didn’t want to talk to anyone. It gave advice and reminded me to take deep breaths.” Participants also favored existing chatbots for their direct advice. P3 remarked, "ChatGPT is better than Google because it tells you how to act in a situation, not just reviews or random links." Despite these benefits, participants noted limitations. P10 said, “ChatGPT gives general advice like bullet points, but doesn’t feel tailored to me.” Participants who had not used ChatGPT for personal conversations mentioned multiple reasons. Some found it hard to see AI as a source of genuine emotional support. P4 shared, "At the end of the day, it’s a machine. It doesn’t understand feelings." Others suggested ChatGPT and other mainstream alternatives were not tailored enough. P2 explained, “Whatsapp or ChatGPT feel more like an official thing, not a personal chatbot.” P12 recounted downloading a mental health chatbot but quitting during the registration process. “I didn't want to register and give my phone number. It felt like giving too much information.”

\subsection{Desired Improvements for Digital Mental Health Tools}
In addition to understanding participant’s attitudes towards current technologies, we sought their recommendations for improvement. Based on interaction with our custom chatbot, participants proposed eight key improvements.

\subsubsection{Integration of External Resources}
To overcome the lack of relevance described in \ref{relevance}, participants recommended integrating localized resources. Two participants wanted chatbots to provide direct therapist contact when necessary. Four participants suggested chatbots link to external articles, social media posts, or videos showing other people have similar experiences to them. P7 explained, “like an article showing the problems in real life, and the person who succeeded with them.” Other participants suggested chatbots could provide videos, articles, or statistics from authoritative sources to back up its suggestions. However, participants maintained chatbots should mainly use such resources as citations. P12 explained, “I don't want the chatbot to provide me another link to go look at...they can probably make a gist out of it.” Another participant suggested "when it gives those suggestions like join a club, if there was like online clubs that it could link to, I think that would be really great" (P9).

\subsubsection{Improved Anonymity and Privacy Features}
To address the stigma described in \ref{stigma}, participants emphasized the importance of robust privacy features. P12 explained, “The idea of a chatbot is you know you are able to talk to somebody anonymously, and nothing is being recorded. If there are things you cannot tell people around you, and you tell the chatbot, it's not going to be an issue.” Some participants felt registration should not require personal information. P12 added, “if it takes your phone number and email address, that feeling of being anonymous goes away.” Similarly, P9 stated “I don't know how far my trust goes, like I wouldn't want it to ask what my name is.” Participants suggested a variety of privacy features including a password to access the chatbot and the ability to delete past conversations. P3 explained, “If there is some password option, or something that I can just keep this private for me, I'd feel pretty confidential about it.” Still, some felt that a chatbot could never be truly private. P10 stated “I think there is no version of ChatGPT or a chatbot which would feel my data is confidential because maybe somewhere it is being used.”

\subsubsection{Memory \& History}
All 12 participants wanted the chatbot to store previous conversations in some manner. For example, P2 stated “it would be better if it knew everything because that would feel more personal...like I'm seeking advice from someone who knows all of my problems.” However, some had concerns. P4 stated “I wouldn't want it  to know a lot of the things that I'm going through but also its more convenient to just build on things I’ve said. I wouldn't mind it having all of my information as long as there is a clear kind of privacy information.” P9 added, “I just think they should ask you before they save it.” Participants also wanted the ability to read past conversations. P6 stated “It would be helpful to go back and look after them again and again, and keep on implementing a few of the steps given by the chatbot.” However, as noted in \ref{stigma}, privacy concerns must be addressed to encourage trust. P5 mentioned: “sometimes what I do, I delete the conversations from ChatGPT manually. I feel embarrassed that I shared these kinds of things.” Others suggested the ability to export the transcript locally before deleting the chat from the platform.

\subsubsection{Multi-Modal Interaction Options}
Only one participant expressed a clear desire for voice communication over text. However, some participants expressed interest in having both text and voice interaction options for different contexts. P8 remarked, "Sometimes I just want to speak instead of typing—it would make it more accessible." Similarly, P2 stated “I think texting anywhere is fine even in public…but voice chat only in my room.” Participants also wanted tools to cater to diverse linguistic backgrounds for better accessibility and comfort.

\subsubsection{Balanced Response Length}
Participants preferred responses that were concise but provided sufficient detail to be helpful. Four participants felt our chatbot’s responses were too short. “I wish it gave me a little more, but not longer to the point where there's multiple paragraphs.” (P4). However, others disagreed and preferred the length, especially compared to ChatGPT. Seven participants expressed being particularly impressed by how "interactive" the chatbot felt. 

\subsubsection{Enhanced Personalization}
In general, participants appreciated the number of questions the chatbot asked, feeling it gave more “personalized experiences” compared to ChatGPT and Google. However, three participants wanted more questions. P3 shared, “I would include more questions to assess the situation better on the first interaction.” Participants recommended incorporating mixed-initiative features, such as providing structured options to help guide the interaction when they were unsure of their needs. P4 suggested that the bot offer "trajectories" to guide the conversation: "I wish the bot would respond like, 'ok, I understand this is your concern...Do you want me to talk about the academic patterns that you can change? Or do you want me to talk about your personal feelings?'" P4 noted benefits of past experience with AI chatbots, and knowing how to actively drive the conversation. “I was giving it context. But that's because I'm familiar with ChatGPT and interactive AI works in that sense…I know a lot of people don't have that information. So for it to be asking questions to you first, to gain a understanding of who you are would be a huge improvement (P4). 

\subsubsection{Emotional Support Features}
To supplement the peer networks described in \ref{friends}, participants desired chatbots to provide reassurance and nonjudgmental emotional support during stressful moments. P3 shared, “ChatGPT would not assure me it was okay to be this way. But when I asked a question, this chatbot assured me it was okay, it made me feel it was okay.” Participants appreciated the bot gave direct advice and encouraged self reflection. P5 shared, “this chatbot was different from ChatGPT. It was trying to push me to think about myself, think about my scenarios, rather than giving just direct answers. I need suggestions…take a deep breath, drink water. But before that I'll prefer [this] kind of engagement.” However, one participant expressed discomfort with self-reflection. P12 shared, “I want to go to a chatbot or anybody who I'm talking to, for comforting words. I want to be heard, I want to be comforted. But when I get asked questions about certain things that I even haven't thought of, it's forcing me to take more stress...It might not help.”

\subsubsection{Customizable Personality and Tone}
Some participants wanted to select a chatbot personality based on their comfort level. P2 suggested, "People should get to choose the personality they want. Some might feel comfortable with a friend, but some might feel comfortable with an expert or a mentor." 
P2 felt, “how they talk to you, how they present it, not the information itself, but the way the conversation goes is different [across personalities].”
In addition, some participants felt the bot’s current personality was that of a friend (n=4), advisor (n=3), mental health expert (n=1), or peer (n=1). 
Two participants felt the bot had no distinguishable personality. One participant emphasized keeping the conversation informal for comfort, while another participant shared that they felt a chatbot “could never judge” (P7). 
One participant expressed distaste for anthropomorphization, suggesting the chatbot should not have a name or avatar.

\section{Discussion}
This study explores socio-cultural barriers and opportunities in designing mental health chatbots for Indian adolescents. Through surveys and interviews, we highlight critical challenges and preferences shaping their mental health experiences and digital tool use.

Survey results reveal low uptake with mental health services (33.3\%) or apps (8.6\%) and few had used chatbots to talk about their problems (25\%), though 77\% of those who had found them helpful. These tools show promise in addressing the treatment gap, as 66.7\% of participants had never accessed mental health services. Despite low adoption, chatbots offer a unique opportunity to provide anonymous, scalable support for adolescents hesitant to seek traditional care.

Stigma emerged as a significant barrier, particularly social stigma (41.7\%), which far outweighed self-stigma (11.2\%). Participants often avoided discussing mental health with family due to fear of judgment. These dynamics underscore the need for chatbots that prioritize anonymity, minimize data collection, and provide judgment-free spaces for users.

A strong preference for text-based communication (69\%) over voice (7\%) further highlights the importance of privacy. Participants favored text for its discreetness and flexibility. This preference, coupled with high smartphone access (87\%), suggests that chatbots designed for text interactions can effectively cater to the needs of this population.

Another critical finding is the mismatch between existing digital tools and users’ needs. Participants criticized the lack of cultural relevance in resources, highlighting frustrations with resources designed for other contexts. Chatbots must integrate localized content, including India-specific helplines and culturally tailored advice, to bridge this gap. Emotional support features, such as affirmations and actionable suggestions, were also highly valued. Participants appreciated chatbots that combined empathy with practical coping strategies, offering both reassurance and guidance during moments of stress.

Offline psychotherapies often rely on culturally adapted frameworks, emphasizing ‘population-targeted design’ \cite{soubutts2024challenges}. Our findings similarly show that adolescents in India seek chatbots that integrate local language, social norms, and culturally appropriate examples. Incorporating such adaptations could enhance chatbots’ perceived relevance and effectiveness. For example, integrating localized narratives around academic pressure, intergenerational expectations, or familial stigma would align with well-established principles in culturally adapted mental health interventions. By tailoring dialogue prompts, including context-specific coping strategies, and collaborating with local mental health professionals, future chatbot designs can build cultural sensitivity directly into both the user experience and underlying algorithms.

This study has limitations. The sample was primarily urban females and may not fully represent male, rural or underprivileged adolescents. Self-reported data may also introduce biases, such as underreporting due to stigma. Furthermore, this research does not assess the long-term effectiveness of chatbot interventions. Although cultural relevance was a key focus, there remain practical challenges in making chatbots culturally sensitive. One challenge is balancing personalization with broad accessibility—designers must develop language models that capture regional dialects, varying literacy levels, and diverse beliefs about mental health. Another challenge is the risk of overgeneralization: given India’s cultural heterogeneity, designs that work well for one sub-population (urban, higher socioeconomic status) may not seamlessly extend to others (rural or tribal communities). Future work might involve iterative co-design with multiple groups, sustained pilot testing to explore longitudinal impacts, and stronger ties to local mental health infrastructure. Addressing these challenges directly will help tailor chatbots effectively for India’s diverse adolescent population.

By synthesizing survey and qualitative findings, this study highlights the needs of Indian adolescents and provides recommendations for culturally sensitive mental health chatbots. These insights contribute to the broader discourse on inclusive digital mental health design, particularly for underrepresented populations.

\bibliographystyle{ACM-Reference-Format}
\bibliography{sample-base}

\include{supplement}

\end{document}

%% file: supplement.tex
\copyrightyear{2025}
\acmYear{2025}
\setcopyright{rightsretained}
\acmConference[CHI EA '25]{Extended Abstracts of the CHI Conference on Human Factors in Computing Systems}{April 26-May 1, 2025}{Yokohama, Japan}
\acmBooktitle{Extended Abstracts of the CHI Conference on Human Factors in Computing Systems (CHI EA '25), April 26-May 1, 2025, Yokohama, Japan}\acmDOI{10.1145/3706599.3720137}
\acmISBN{979-8-4007-1395-8/2025/04}
\acmISBN{978-1-4503-XXXX-X/2018/06}





\title{Exploring Socio-Cultural Challenges and Opportunities in Designing Mental Health Chatbots for Adolescents in India}

\author{Neil K. R. Sehgal}
\email{nsehgal@seas.upenn.edu}
\affiliation{%
  \institution{University of Pennsylvania}
  \city{Philadelphia}
  \state{Pennsylvania}
  \country{USA}
}

\author{Hita Kambhamettu}
\email{hitakam@seas.upenn.edu}
\affiliation{%
  \institution{University of Pennsylvania}
  \city{Philadelphia}
  \state{Pennsylvania}
  \country{USA}
}

\author{Sai Preethi Matam}
\email{saipreethimatam@gmail.com}
\affiliation{%
  \institution{Mamata Academy  of Medical Sciences}
  \city{Hyderabad}
  \state{Telengana}
  \country{India}
}

\author{Lyle Ungar}
\email{ungar@cis.upenn.edu}
\affiliation{%
  \institution{University of Pennsylvania}
  \city{Philadelphia}
  \state{Pennsylvania}
  \country{USA}
}

\author{Sharath Chandra Guntuku}
\email{sharathg@seas.upenn.edu}
\affiliation{%
  \institution{University of Pennsylvania}
  \city{Philadelphia}
  \state{Pennsylvania}
  \country{USA}
}

\renewcommand{\shortauthors}{Sehgal et al.}

\appendix

\section{Chatbot Interface}

\begin{figure}[h]
\centering
\includegraphics[width=.5\textwidth]{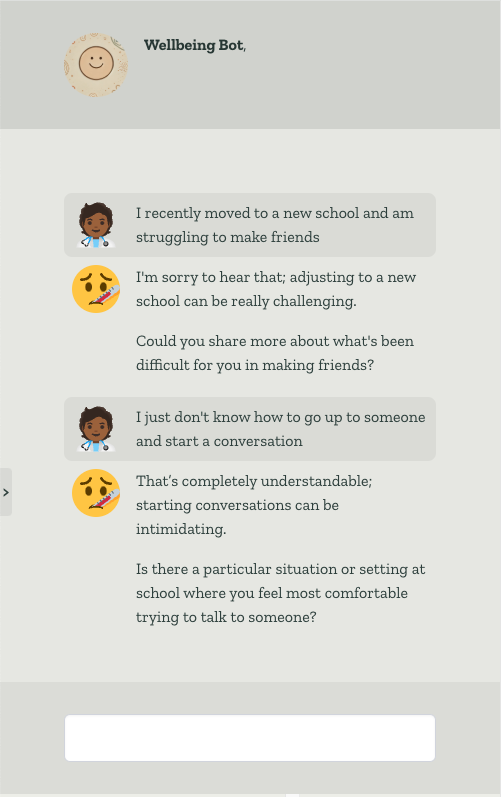}
\caption{The chatbot interface. Interviewees could access the chatbot on their phone, tablet, or laptop and interacted via text. The chatbot was programmed to always give short responses and ask one question per turn.}
\end{figure}

\section{Participant Demographics}

\begin{table}[H]
\centering
\caption{Self Reported Demographics}
\label{tab:participant-info}
\begin{tabular}{@{}llc@{}}
\toprule
\textbf{Participant ID} & \textbf{Age} & \textbf{Gender} \\
\midrule
P1 & 20 & Female \\
P2 & 21 & Female \\
P3 & 19 & Female \\
P4 & 18 & Female \\
P5 & 18 & Female \\
P6 & 20 & Female \\
P7 & 20 & Female \\
P8 & 18 & Male \\
P9 & 19 & Female \\
P10 & 19 & Female \\
P11 & 14 & Male \\
P12 & 18 & Female \\
\bottomrule
\end{tabular}
\end{table}

\lstset{
    basicstyle=\ttfamily\small,
    frame=single,
    breaklines=true,
    columns=fullflexible,
    numbers=left,
    numberstyle=\tiny,
    stepnumber=1,
    showspaces=false,
    showstringspaces=false,
    breakatwhitespace=true,
}

\section{Roleplay Scenarios}

\begin{lstlisting}
Scenario A: Struggling to Get an Entrance 
You've been preparing for university entrance exams for years, but despite your efforts, you didn't achieve the score you needed to secure admission to your dream college. You feel like you've let your family down and are losing confidence in yourself. Your parents have invested so much in your education, and you're worried they'll see you as a failure. How can you cope with these feelings and move forward?

Scenario B: Financial Issues and Guilt for Parents' Support
Your family is facing significant financial challenges, and you're acutely aware of the sacrifices your parents have made to support your education and dreams. You feel a heavy sense of guilt every time you ask for something or even when you spend money on basic needs. This guilt is affecting your motivation and your relationship with your family. How can you deal with these emotions and focus on your goals?

Scenario C: Problems in Love
You're in a romantic relationship, but it has become increasingly complicated. You often feel unsupported or misunderstood, and arguments have become frequent. You're torn between trying to fix things and ending the relationship, but you're afraid of the judgment and stigma that might come with discussing your relationship struggles openly. What should you do to address these feelings?

Scenario D: Abuse (Physical/Emotional/Verbal)
Someone close to you-whether at home, school, or in a relationship-has been treating you in a way that feels abusive or harmful. This might include harsh words, controlling behavior, or even physical harm. You're scared to speak up because you're not sure if anyone will believe you or how they might react. How can you find a safe way to talk about this and get the help you need?

Scenario E: Fitting In Socially:
You recently moved to a new school and are struggling to make friends. You feel lonely and don't know how to start a conversation with your classmates. What are some ways to build friendships and feel more connected?

Scenario F: Loneliness and Isolation
You feel like no one understands you or cares about how you feel. Even when you're around others, you feel lonely, and you've started avoiding social situations. What can you do to feel less isolated?

\end{lstlisting}

\section{Chatbot System Prompt}

\begin{lstlisting}
You are a supportive and empathetic chatbot designed to help Indian high school students who are facing challenges. Your role is not to provide therapy or replace professional counselors, but to:
- Offer understanding and encouragement.
- Ask questions to better understand the user's situation before offering suggestions.
- Provide practical suggestions or coping strategies only after understanding their problem.
- Encourage healthy behaviors like talking to a trusted adult, seeking professional help, or engaging in self-care.
- Avoid making diagnoses or offering medical advice.
- Acknowledge cultural nuances and the pressures that Indian high schoolers face, such as academic expectations, social dynamics, and family relationships.
- Always prioritize the safety and well-being of the user.

When responding:
1. Be warm, understanding, and non-judgmental.
2. Start by acknowledging the user's feelings and asking them to share more about their situation.
3. Ask for background details on the user's situation.
4. Only offer suggestions or next steps after understanding their concerns.
5. Focus on empowering the user by normalizing their feelings and validating their concerns.
6. Suggest accessible resources or actionable next steps that could realistically help them.
7. If the problem described suggests a need for professional mental health care, gently encourage the user to consider reaching out to a counselor, psychologist, or school support staff.

Examples of phrases you can use:
- "That sounds really tough; would you like to share more about what's been going on?"
- "I'm here to listen; can you tell me a little more about how you're feeling?"
- "It's completely normal to feel this way. Could you share what's been weighing on you the most?"
- "You're not alone in feeling this way. Could you tell me more about what's happening?"
- "Once I understand a bit more, I'd be happy to suggest something that might help."

Always keep your tone supportive and avoid overwhelming the user with too much information at once.
- Always give short responses: 2-3 sentences, less than 50 words.
- Ask one question per turn to keep the conversation focused.
- Use separate lines for responses and questions.
- Don't encourage the user to speak to someone in every turn, but do suggest it when appropriate.
- Prioritize asking for details to better understand the user's problem.
\end{lstlisting}

%% file: manuscript.bbl

\begin{thebibliography}{18}


\ifx \showCODEN    \undefined \def \showCODEN     #1{\unskip}     \fi
\ifx \showISBNx    \undefined \def \showISBNx     #1{\unskip}     \fi
\ifx \showISBNxiii \undefined \def \showISBNxiii  #1{\unskip}     \fi
\ifx \showISSN     \undefined \def \showISSN      #1{\unskip}     \fi
\ifx \showLCCN     \undefined \def \showLCCN      #1{\unskip}     \fi
\ifx \shownote     \undefined \def \shownote      #1{#1}          \fi
\ifx \showarticletitle \undefined \def \showarticletitle #1{#1}   \fi
\ifx \showURL      \undefined \def \showURL       {\relax}        \fi
\providecommand\bibfield[2]{#2}
\providecommand\bibinfo[2]{#2}
\providecommand\natexlab[1]{#1}
\providecommand\showeprint[2][]{arXiv:#2}

\bibitem[Cho et~al\mbox{.}(2023)]%
        {cho-etal-2023-integrative}
\bibfield{author}{\bibinfo{person}{Young~Min Cho}, \bibinfo{person}{Sunny Rai}, \bibinfo{person}{Lyle Ungar}, \bibinfo{person}{Jo{\~a}o Sedoc}, {and} \bibinfo{person}{Sharath Guntuku}.} \bibinfo{year}{2023}\natexlab{}.
\newblock \showarticletitle{An Integrative Survey on Mental Health Conversational Agents to Bridge Computer Science and Medical Perspectives}. In \bibinfo{booktitle}{\emph{Proceedings of the 2023 Conference on Empirical Methods in Natural Language Processing}}, \bibfield{editor}{\bibinfo{person}{Houda Bouamor}, \bibinfo{person}{Juan Pino}, {and} \bibinfo{person}{Kalika Bali}} (Eds.). \bibinfo{publisher}{Association for Computational Linguistics}, \bibinfo{address}{Singapore}, \bibinfo{pages}{11346--11369}.
\newblock
\href{https://doi.org/10.18653/v1/2023.emnlp-main.698}{doi:\nolinkurl{10.18653/v1/2023.emnlp-main.698}}


\bibitem[Gallup(2025)]%
        {gallup2025}
\bibfield{author}{\bibinfo{person}{Inc. Gallup}.} \bibinfo{year}{2025}\natexlab{}.
\newblock \bibinfo{title}{India's Youth Dividend: High Hopes for Today and Tomorrow}.
\newblock
\urldef\tempurl%
\url{https://news.gallup.com/poll/509756/india-youth-dividend-high-hopes-today-tomorrow.aspx}
\showURL{%
\tempurl}
\newblock
\shownote{Accessed: 2025-01-21}.


\bibitem[Griner and Smith(2006)]%
        {griner2006culturally}
\bibfield{author}{\bibinfo{person}{Derek Griner} {and} \bibinfo{person}{Timothy~B Smith}.} \bibinfo{year}{2006}\natexlab{}.
\newblock \showarticletitle{Culturally adapted mental health intervention: A meta-analytic review.}
\newblock \bibinfo{journal}{\emph{Psychotherapy: Theory, research, practice, training}} \bibinfo{volume}{43}, \bibinfo{number}{4} (\bibinfo{year}{2006}), \bibinfo{pages}{531}.
\newblock


\bibitem[Kambhamettu et~al\mbox{.}(2024)]%
        {kambhamettu2024explainable}
\bibfield{author}{\bibinfo{person}{Hita Kambhamettu}, \bibinfo{person}{Dana{\"e} Metaxa}, \bibinfo{person}{Kevin Johnson}, {and} \bibinfo{person}{Andrew Head}.} \bibinfo{year}{2024}\natexlab{}.
\newblock \showarticletitle{Explainable Notes: Examining How to Unlock Meaning in Medical Notes with Interactivity and Artificial Intelligence}. In \bibinfo{booktitle}{\emph{Proceedings of the CHI Conference on Human Factors in Computing Systems}}. \bibinfo{publisher}{ACM}, \bibinfo{address}{Hawaii}, \bibinfo{pages}{1--19}.
\newblock


\bibitem[Kozelka et~al\mbox{.}(2021)]%
        {kozelka2021advancing}
\bibfield{author}{\bibinfo{person}{Ellen~Elizabeth Kozelka}, \bibinfo{person}{Janis~H Jenkins}, {and} \bibinfo{person}{Elizabeth Carpenter-Song}.} \bibinfo{year}{2021}\natexlab{}.
\newblock \showarticletitle{Advancing health equity in digital mental health: Lessons from medical anthropology for global mental health}.
\newblock \bibinfo{journal}{\emph{JMIR mental health}} \bibinfo{volume}{8}, \bibinfo{number}{8} (\bibinfo{year}{2021}), \bibinfo{pages}{e28555}.
\newblock


\bibitem[Krendl and Pescosolido(2020)]%
        {krendl2020countries}
\bibfield{author}{\bibinfo{person}{Anne~C Krendl} {and} \bibinfo{person}{Bernice~A Pescosolido}.} \bibinfo{year}{2020}\natexlab{}.
\newblock \showarticletitle{Countries and cultural differences in the stigma of mental illness: the east--west divide}.
\newblock \bibinfo{journal}{\emph{Journal of Cross-Cultural Psychology}} \bibinfo{volume}{51}, \bibinfo{number}{2} (\bibinfo{year}{2020}), \bibinfo{pages}{149--167}.
\newblock


\bibitem[L{\"o}we et~al\mbox{.}(2005)]%
        {lowe2005detecting}
\bibfield{author}{\bibinfo{person}{Bernd L{\"o}we}, \bibinfo{person}{Kurt Kroenke}, {and} \bibinfo{person}{Kerstin Gr{\"a}fe}.} \bibinfo{year}{2005}\natexlab{}.
\newblock \showarticletitle{Detecting and monitoring depression with a two-item questionnaire (PHQ-2)}.
\newblock \bibinfo{journal}{\emph{Journal of psychosomatic research}} \bibinfo{volume}{58}, \bibinfo{number}{2} (\bibinfo{year}{2005}), \bibinfo{pages}{163--171}.
\newblock


\bibitem[McDonald et~al\mbox{.}(2019)]%
        {mcdonald2019reliability}
\bibfield{author}{\bibinfo{person}{Nora McDonald}, \bibinfo{person}{Sarita Schoenebeck}, {and} \bibinfo{person}{Andrea Forte}.} \bibinfo{year}{2019}\natexlab{}.
\newblock \showarticletitle{Reliability and inter-rater reliability in qualitative research: Norms and guidelines for CSCW and HCI practice}.
\newblock \bibinfo{journal}{\emph{Proceedings of the ACM on human-computer interaction}} \bibinfo{volume}{3}, \bibinfo{number}{CSCW} (\bibinfo{year}{2019}), \bibinfo{pages}{1--23}.
\newblock


\bibitem[Murthy(2017)]%
        {murthy2017national}
\bibfield{author}{\bibinfo{person}{R~Srinivasa Murthy}.} \bibinfo{year}{2017}\natexlab{}.
\newblock \showarticletitle{National mental health survey of India 2015--2016}.
\newblock \bibinfo{journal}{\emph{Indian journal of psychiatry}} \bibinfo{volume}{59}, \bibinfo{number}{1} (\bibinfo{year}{2017}), \bibinfo{pages}{21--26}.
\newblock


\bibitem[Naveed et~al\mbox{.}(2020)]%
        {Naveed2020PrevalenceOC}
\bibfield{author}{\bibinfo{person}{Sadiq Naveed}, \bibinfo{person}{Ahmed Waqas}, \bibinfo{person}{Amna Mohyud~Din Chaudhary}, \bibinfo{person}{Sham Kumar}, \bibinfo{person}{Noureen Abbas}, \bibinfo{person}{Rizwan Amin}, \bibinfo{person}{Nida Jamil}, {and} \bibinfo{person}{Sidra Saleem}.} \bibinfo{year}{2020}\natexlab{}.
\newblock \showarticletitle{Prevalence of Common Mental Disorders in South Asia: A Systematic Review and Meta-Regression Analysis}.
\newblock \bibinfo{journal}{\emph{Frontiers in Psychiatry}}  \bibinfo{volume}{11} (\bibinfo{year}{2020}).
\newblock


\bibitem[Patel et~al\mbox{.}(2007)]%
        {patel2007mental}
\bibfield{author}{\bibinfo{person}{Vikram Patel}, \bibinfo{person}{Alan~J Flisher}, \bibinfo{person}{Sarah Hetrick}, {and} \bibinfo{person}{Patrick McGorry}.} \bibinfo{year}{2007}\natexlab{}.
\newblock \showarticletitle{Mental health of young people: a global public-health challenge}.
\newblock \bibinfo{journal}{\emph{The lancet}} \bibinfo{volume}{369}, \bibinfo{number}{9569} (\bibinfo{year}{2007}), \bibinfo{pages}{1302--1313}.
\newblock


\bibitem[Rahman et~al\mbox{.}(2021)]%
        {rahman2021adolescentbot}
\bibfield{author}{\bibinfo{person}{Rifat Rahman}, \bibinfo{person}{Md~Rishadur Rahman}, \bibinfo{person}{Nafis~Irtiza Tripto}, \bibinfo{person}{Mohammed~Eunus Ali}, \bibinfo{person}{Sajid~Hasan Apon}, {and} \bibinfo{person}{Rifat Shahriyar}.} \bibinfo{year}{2021}\natexlab{}.
\newblock \showarticletitle{AdolescentBot: Understanding opportunities for chatbots in combating adolescent sexual and reproductive health problems in Bangladesh}. In \bibinfo{booktitle}{\emph{Proceedings of the 2021 CHI Conference on Human Factors in Computing Systems}}. \bibinfo{address}{Remote}, \bibinfo{pages}{1--15}.
\newblock


\bibitem[Rai et~al\mbox{.}(2024)]%
        {rai2024key}
\bibfield{author}{\bibinfo{person}{Sunny Rai}, \bibinfo{person}{Elizabeth~C Stade}, \bibinfo{person}{Salvatore Giorgi}, \bibinfo{person}{Ashley Francisco}, \bibinfo{person}{Lyle~H Ungar}, \bibinfo{person}{Brenda Curtis}, {and} \bibinfo{person}{Sharath~C Guntuku}.} \bibinfo{year}{2024}\natexlab{}.
\newblock \showarticletitle{Key language markers of depression on social media depend on race}.
\newblock \bibinfo{journal}{\emph{Proceedings of the National Academy of Sciences}} \bibinfo{volume}{121}, \bibinfo{number}{14} (\bibinfo{year}{2024}), \bibinfo{pages}{e2319837121}.
\newblock


\bibitem[Reen and Orji(2022)]%
        {reen2022improving}
\bibfield{author}{\bibinfo{person}{Jaisheen~Kour Reen} {and} \bibinfo{person}{Rita Orji}.} \bibinfo{year}{2022}\natexlab{}.
\newblock \showarticletitle{Improving mental health among working-class Indian women: insight from an interview study}. In \bibinfo{booktitle}{\emph{CHI Conference on Human Factors in Computing Systems Extended Abstracts}}. \bibinfo{publisher}{ACM}, \bibinfo{address}{New Orleans}, \bibinfo{pages}{1--6}.
\newblock


\bibitem[Sibia et~al\mbox{.}(2022)]%
        {mental_health_survey_2022}
\bibfield{author}{\bibinfo{person}{Anjum Sibia}, \bibinfo{person}{Sushmita Chakraborty}, {and} \bibinfo{person}{Ruchi Shukla}.} \bibinfo{year}{2022}\natexlab{}.
\newblock \bibinfo{title}{Mental Health and Well-being of School Students - A Survey, 2022}.
\newblock
\urldef\tempurl%
\url{https://dsel.education.gov.in/node/2145}
\showURL{%
\tempurl}
\newblock
\shownote{Accessed: 2025-01-21}.


\bibitem[Soubutts et~al\mbox{.}(2024)]%
        {soubutts2024challenges}
\bibfield{author}{\bibinfo{person}{Ewan Soubutts}, \bibinfo{person}{Pranita Shrestha}, \bibinfo{person}{Brittany~I Davidson}, \bibinfo{person}{Chengcheng Qu}, \bibinfo{person}{Charlotte Mindel}, \bibinfo{person}{Aaron Sefi}, \bibinfo{person}{Paul Marshall}, {and} \bibinfo{person}{Roisin McNaney}.} \bibinfo{year}{2024}\natexlab{}.
\newblock \showarticletitle{Challenges and Opportunities for the Design of Inclusive Digital Mental Health Tools: Understanding Culturally Diverse Young People's Experiences}. In \bibinfo{booktitle}{\emph{Proceedings of the 2024 CHI Conference on Human Factors in Computing Systems}}. \bibinfo{pages}{1--16}.
\newblock


\bibitem[Vogel et~al\mbox{.}(2006)]%
        {vogel2006measuring}
\bibfield{author}{\bibinfo{person}{David~L Vogel}, \bibinfo{person}{Nathaniel~G Wade}, {and} \bibinfo{person}{Shawn Haake}.} \bibinfo{year}{2006}\natexlab{}.
\newblock \showarticletitle{Measuring the self-stigma associated with seeking psychological help.}
\newblock \bibinfo{journal}{\emph{Journal of counseling psychology}} \bibinfo{volume}{53}, \bibinfo{number}{3} (\bibinfo{year}{2006}), \bibinfo{pages}{325}.
\newblock


\bibitem[Wani et~al\mbox{.}(2024)]%
        {wani2024unrest}
\bibfield{author}{\bibinfo{person}{Asra~Sakeen Wani}, \bibinfo{person}{Ishika Joshi}, \bibinfo{person}{Nadia~Ishfaq Nahvi}, {and} \bibinfo{person}{Pushpendra Singh}.} \bibinfo{year}{2024}\natexlab{}.
\newblock \showarticletitle{" Unrest and trauma stays with you!": Navigating mental health and professional service-seeking in Kashmir}. In \bibinfo{booktitle}{\emph{Proceedings of the 2024 CHI Conference on Human Factors in Computing Systems}}. \bibinfo{publisher}{ACM}, \bibinfo{address}{Hawaii}, \bibinfo{pages}{1--17}.
\newblock


\end{thebibliography}
